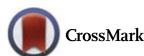

# New Journal of Physics
The open access journal at the forefront of physics

Deutsche Physikalische Gesellschaft **DPG**

**IOP** Institute of Physics

Published in partnership with: Deutsche Physikalische Gesellschaft and the Institute of Physics

CrossMark

**OPEN ACCESS**

PAPER

# Monte Carlo sampling from the quantum state space. II

Yi-Lin Seah[1,2], Jiangwei Shang[1], Hui Khoon Ng[1,3,4], David John Nott[5] and Berthold-Georg Englert[1,2,4]

[1] Centre for Quantum Technologies, National University of Singapore, 3 Science Drive 2, Singapore 117543, Singapore
[2] Department of Physics, National University of Singapore, 2 Science Drive 3, Singapore 117542, Singapore
[3] Yale-NUS College, 6 College Avenue East, Singapore 138614, Singapore
[4] MajuLab, CNRS-UNS-NUS-NTU International Joint Research Unit, UMI 3654, Singapore
[5] Department of Statistics and Applied Probability, National University of Singapore, 6 Science Drive 2, Singapore 117546, Singapore

E-mail: s89@u.nus.edu and jiangwei.shang@quantumlah.org

Keywords: Monte Carlo, sampling, quantum state space, Hamiltonian Monte Carlo, hybrid Monte Carlo

## Abstract

High-quality random samples of quantum states are needed for a variety of tasks in quantum information and quantum computation. Searching the high-dimensional quantum state space for a global maximum of an objective function with many local maxima or evaluating an integral over a region in the quantum state space are but two exemplary applications of many. These tasks can only be performed reliably and efficiently with Monte Carlo methods, which involve good samplings of the parameter space in accordance with the relevant target distribution. We show how the Markov-chain Monte Carlo method known as Hamiltonian Monte Carlo, or hybrid Monte Carlo, can be adapted to this context. It is applicable when an efficient parameterization of the state space is available. The resulting random walk is entirely inside the physical parameter space, and the Hamiltonian dynamics enable us to take big steps, thereby avoiding strong correlations between successive sample points while enjoying a high acceptance rate. We use examples of single and double qubit measurements for illustration.

## 1. Introduction

Our companion paper [1] states the motivation for this work and introduces the terminology and notational conventions we are using; when referring to an equation or figure in that paper, the respective number is preceded by 'I-'. While the sampling methods presented there—rejection sampling, importance sampling, and Markov-chain sampling—are easy to implement, they require a costly (in CPU time) check whether the candidate probabilities obey all constraints imposed by the positivity of the statistical operator. By contrast, in the Hamiltonian Monte Carlo (HMC) method [2, 3] that we discuss here, the constraints are always obeyed by construction.

    Like the Markov-chain Monte Carlo (MCMC) method discussed in I, HMC involves a 'walk' around the parameter space. While one often needs a walk comprising many small steps for MCMC to attain a good sampling yield, HMC can be done with large steps. As a consequence, the problem of a slow exploration of the probability space, in conjunction with strong correlations between successive sample points, which is a typical feature of other random-walk strategies, is not present in HMC.

    Further, HMC enables us to sample in accordance with any (reasonable) prior density and any posterior density. Some other sampling methods are very efficient for particular priors (see, e.g., [4, 5]), but lack this much-needed flexibility. The downside of HMC, however, is the requirement for a suitable parameterization for its implementation.

## 2. Hamiltonian Monte Carlo

### 2.1. The idea

The HMC method makes use of pseudo-Hamiltonian dynamics in a mock phase space and can be applied to most problems with continuous state spaces by introducing fictitious momentum variables. One way of conveying the basic idea of HMC is the following.

We begin with a trial density $f(\theta)$, supplement the position variables $\theta = (\theta_1, \theta_2, ..., \theta_S)$ with momentum variables $\vartheta = (\vartheta_1, \vartheta_2, ..., \vartheta_S)$, and dress up the position density $f(\theta)$ with a Gaussian momentum density to compose the initial phase-space density

$$F_0(\theta, \vartheta) = f(\theta) \prod_{s=1}^{S} \frac{1}{\sqrt{2\pi}} e^{-\frac{1}{2}\vartheta_s^2} = (2\pi)^{-S/2} f(\theta) e^{-\frac{1}{2}\sum_s \vartheta_s^2} \qquad (1)$$

at time $t = 0$. We obtain the final phase-space density at time $t = T$ by propagating $F_0$ to $F_T$ with the aid of Liouville's equation

$$\frac{\partial}{\partial t} F_t(\theta, \vartheta) = -\mathcal{D} F_t(\theta, \vartheta), \qquad (2)$$

where the differential operator

$$\mathcal{D} = \sum_s \left( \frac{\partial H}{\partial \vartheta_s} \frac{\partial}{\partial \theta_s} - \frac{\partial H}{\partial \theta_s} \frac{\partial}{\partial \vartheta_s} \right) \qquad (3)$$

involves the velocity components $\frac{\partial}{\partial \vartheta_s} H = \vartheta_s$ and the force components $u_s(\theta) = -\frac{\partial}{\partial \theta_s} H = w(\theta)^{-1} \frac{\partial}{\partial \theta_s} w(\theta)$ associated with the Hamiltonian

$$H(\theta, \vartheta) = \frac{1}{2} \sum_s \vartheta_s^2 - \log w(\theta). \qquad (4)$$

At the final time $T$, we thus have

$$F_T(\theta, \vartheta) = e^{-T\mathcal{D}} F_0(\theta, \vartheta). \qquad (5)$$

The updated position density $\tilde{f}(\theta)$ now results from integrating over the momenta, so that

$$f(\theta) \to \tilde{f}(\theta) = \int \frac{(d\vartheta)}{(2\pi)^{S/2}} e^{-T\mathcal{D}} f(\theta) e^{-\frac{1}{2}\sum_s \vartheta_s^2} \qquad (6)$$

is the net map for $f(\theta)$. Exceptional situations aside (usually they arise from an unfortunate choice for the duration $T$), $f(\theta)$ is a fixed point of this map only if $F_T = F_0$ is a function of the Hamiltonian, which is the case if

$$f(\theta) \doteq e^{\log w(\theta)} = w(\theta), \qquad (7)$$

where the dotted equal sign stands for 'equal up to a constant numerical factor'. Taking for granted without proof that the map (6) is a contraction, repeated applications of the map will thus turn $f(\theta)$ into $w(\theta)$, the desired prior or posterior density.

### 2.2. The implementation

While this conveys the idea of HMC, its actual implementation is, however, not in terms of a trial sample that is iteratively updated by the mapping (6), but by a Markovian random walk. At each step of the walk, the state follows a trajectory $(\theta(t), \vartheta(t))$ from the current position $\theta = \theta(t = 0)$ to the proposal $\theta^\star = \theta(t = T)$, where the components of the initial momentum $\vartheta(t = 0)$ are chosen at random from the Gaussian distribution $\propto e^{-\frac{1}{2}\sum_s \vartheta_s^2}$, as required by the initial phase-space density in (1). As long as the map (6) is ergodic, the time averaged distribution of $\theta$ will converge towards the stationary distribution, with density $w(\theta)$ as demonstrated in section 2.1.

Here, then, is the HMC algorithm [3]:

**HMC1** Start at $j = 1$ with an arbitrary initial point $\theta^{(1)}$.

**HMC2** Generate $\vartheta^{(j)}$ from a multivariate normal distribution with unit variance.

**HMC3** Solve the Hamilton equations of motion

$$\frac{d}{dt} \theta_s(t) = \vartheta_s(t), \qquad \frac{d}{dt} \vartheta_s(t) = u_s(\theta(t)) \qquad (8)$$

for the initial conditions $(\theta, \vartheta)|_{t=0} = \left(\theta^{(j)}, \vartheta^{(j)}\right)$ and obtain $(\theta^\star, \vartheta^\star) = (\theta, -\vartheta)|_{t=T}$.

**HMC4** Calculate the acceptance ratio

$$a = \min\left\{e^{H(\theta^{(j)},\vartheta^{(j)})-H(\theta^\star,\vartheta^\star)}, 1\right\}. \tag{9}$$

**HMC5** Draw a random number $b$ uniformly from the range $0 < b < 1$. Set $\theta^{(j+1)} = \theta^\star$ if $a > b$; otherwise, set $\theta^{(j+1)} = \theta^{(j)}$.

**HMC6** Update $j \to j+1$. Escape the loop when $j = M$, the target number of sample points; otherwise, return to HMC2.

A few remarks are in order. First, the value of the Hamiltonian is constant along the *exact* trajectory $(\theta(t), \vartheta(t))$ of HMC3, which would give $a = 1$ in HMC4. In practice, however, we rely on an approximate trajectory; accepting Neal's advice [3], we calculate it with the leapfrog method described in appendix, and the difference between the initial and final values of $H(\theta, \vartheta)$ in (9) is nonzero.

Second, HMC is an implementation of the Metropolis–Hastings Monte Carlo (MHMC) algorithm [6] that, regardless of the step size used, achieves a high acceptance rate with much weaker correlations between successive points. In MHMC, the proposal $\theta^\star$ is accepted with probability (see (I-D.1)),

$$a = \min\left\{\frac{w(\theta^\star)J(\theta|\theta^\star)}{w(\theta)J(\theta^\star|\theta)}, 1\right\}, \tag{10}$$

where $w(\theta)$ is the target probability density, and $J(\theta^\star|\theta)$ is the probability of proposing point $\theta^\star$ given the current point $\theta$. The comparison with (9) establishes that

$$\frac{J(\theta|\theta^\star)}{J(\theta^\star|\theta)} = \exp\left(\frac{1}{2}\sum_s \left(\vartheta_s^2 - \vartheta_s^{\star 2}\right)\right) \tag{11}$$

in HMC where $\vartheta$ is the randomly chosen initial momentum and $\vartheta^\star$ is the negative of the resulting final momentum; the otherwise irrelevant minus sign in $\vartheta^\star = -\vartheta(t = T)$ in HMC3 ensures the symmetry of the process, thereby exploiting the time-reversal invariance of the equations of motion (8). In effect, then, HMC achieves an acceptance rate close to 1 by a proposal scheme where $\frac{J(\theta|\theta^\star)}{J(\theta^\star|\theta)}$ is close to $\frac{w(\theta)}{w(\theta^\star)}$.

Third, rather than the 'kinetic energy' $\frac{1}{2}\sum_s \vartheta_s^2$ of (4), we could use a general quadratic form $\frac{1}{2}\sum_{s,s'}\vartheta_s\mu_{s,s'}\vartheta_{s'}$ where the coefficients $\mu_{s,s'}$ are the entries of a symmetric positive-definite $S \times S$ matrix; HMC2 and the velocities in HMC3 are then changed accordingly. Whether or not this freedom in choosing matrix $\mu$ offers an advantage, depends on the structure of the 'potential energy' $-\log w(\theta)$. One would usually carry over any symmetries in the potential energy to the kinetic energy; for instance, if $w(\theta)$ is invariant under the interchange $\theta_1 \leftrightarrow \theta_2$, the kinetic energy should treat $\theta_1$ and $\theta_2$ on equal footing. The kinetic energy of (4) is used for all the examples in section 4.

## 3. State parameterization

In [1], all sampling is done in the probability space. Unless the circumstances are so simple that we can state explicitly all constraints obeyed by the probabilities and do not need to execute a physicality check of the kind discussed in appendix A of [1], it is not feasible to implement the HMC random walk for variables $\theta$ that are (a non-redundant subset of the) probabilities. Rather, we parameterize the statistical operator $\rho$, and the $\theta$ dependence of the probabilities then follows from the Born rule.

For a $d$-level quantum system, the statistical operator $\rho$ is represented by a hermitian unit-trace $d \times d$ matrix that has $S = d^2 - 1$ real parameters. The matrix of the arbitrary operator $A$ in $\rho = A^\dagger A/\text{tr}\{A^\dagger A\}$, however, has $2d^2$ real parameters, of which $d^2 + 1$ are superfluous. We get rid of them by restricting the $A$ matrix to upper-triangular (or lower-triangular, as in [7]) form with real diagonal elements; this reduces the count of real parameters to $d^2$. One more parameter is removed by enforcing that

$$\rho = A^\dagger A \text{ and hence } \text{tr}\{A^\dagger A\} = \sum_{1 \leq j \leq k \leq d}|A_{jk}|^2 = 1 \tag{12}$$

holds, which requires that the moduli of the elements of $A$ are points on a $\left(\frac{1}{2}d(d+1) - 1\right)$-sphere. We parameterize this sphere with spherical coordinates—angle parameters $\theta_1, \theta_2, \ldots, \theta_{\frac{1}{2}(d+2)(d-1)}$—with the Cartesian coordinates $C_1, C_2, \ldots, C_{\frac{1}{2}(d+2)(d-1)}, S_{\frac{1}{2}(d+2)(d-1)}$ recursively defined by

$$C_1 = \cos\theta_1, \ S_1 = \sin\theta_1;$$
$$C_k = S_{k-1}\cos\theta_k, \ S_k = S_{k-1}\sin\theta_k \text{ for } k = 2, 3,..., \tfrac{1}{2}(d+2)(d-1). \tag{13}$$

We fill up the upper-triangle of matrix $A$ with the Cartesian coordinates, which are $\tfrac{1}{2}(d+2)(d-1) + 1 = \tfrac{1}{2}d(d+1)$ in number, and supplement the off-diagonal entries with phase factors

$$E_k = e^{-i\theta_k} \quad \text{with } k = \tfrac{1}{2}d(d+1),...,d^2 - 1. \tag{14}$$

The cases $d = 2$, $d = 3$, and $d = 4$ illustrate the matter:

$$\begin{pmatrix} C_1 & C_2 E_3 \\ 0 & S_2 \end{pmatrix}, \quad \begin{pmatrix} C_1 & C_2 E_6 & C_4 E_7 \\ 0 & C_3 & C_5 E_8 \\ 0 & 0 & S_5 \end{pmatrix}, \quad \begin{pmatrix} C_1 & C_2 E_{10} & C_4 E_{11} & C_7 E_{13} \\ 0 & C_3 & C_5 E_{12} & C_8 E_{14} \\ 0 & 0 & C_6 & C_9 E_{15} \\ 0 & 0 & 0 & S_9 \end{pmatrix}. \tag{15}$$

Other ways of assigning the Cartesian coordinates to the upper-triangle entries and the phase factors to the off-diagonal entries are, of course, possible and give equally valid parameterizations. The assignment chosen is

$$A_{jk} = \begin{cases} 0, & \text{if } j > k, \\ C_{\tfrac{1}{2}k(k+1)}, & \text{if } j = k < d, \\ S_{\tfrac{1}{2}(d+2)(d-1)}, & \text{if } j = k = d, \\ C_m E_{m+n}, & \text{if } j < k, \end{cases} \tag{16}$$

with

$$m = \tfrac{1}{2}k(k-1) + j, \quad n = \tfrac{1}{2}(d+2)(d-1) - (k-1). \tag{17}$$

Upon expressing the probabilities $p = (p_1,...,p_K)$ in terms of the parameters $\theta = (\theta_1,...,\theta_S)$ with $S = d^2 - 1$, the step-function constraints in $w_{\text{cstr}}(p)$ of (I-6) are taken care of. Then, integrating out the delta-function constraints gets rid of redundant probability variables. Finally, we need the Jacobian between the remaining $p_k$s and the $\theta_s$s to convert the prior or posterior density in $p$ into the corresponding expression in $\theta$

$$w(p) \to w(\theta) \equiv \left(w(p) \left|\frac{\partial p}{\partial \theta}\right|\right)\bigg|_{p \text{ in terms of } \theta}. \tag{18}$$

The HMC algorithm can now be executed for $u_s(\theta) = \frac{\partial}{\partial \theta_s} \log w(\theta)$.

As stated, the parameterization of (12) and (16) with (13), (14), and (17) is applicable to POMs that are informationally complete, for which the whole state space is the reconstruction space. If the POM is not informationally complete, it may still be possible to use $A$ of (16) with fixed values for some of the $\theta_s$s. An example is the trine measurement for a qubit, for which the $2 \times 2$ matrix in (15) with $\theta_1 = \frac{\pi}{4}$ does the job; see section 4.2. If no such restricted version of (16) serves the purpose, it may be possible to introduce additional parameters during the HMC sampling, thus produce a sample in a space of higher dimension, then marginalize the auxiliary parameters, and so arrive at a proper sample; see section 4.3 for an example.

## 4. Examples

The parameterization of first the statistical operator $\rho$ and then the probabilities $p$ in terms of the angles $\theta$, while systematic, tends to be involved and does not lend itself to simplification when explicit expressions are needed. Therefore, we shall only present detailed examples for the $d = 2$ case of a qubit (sections 4.1 and 4.2), while providing an example with incomplete tomographic data for the $d = 4$ case of a qubit pair (section 4.3).

### 4.1. Qubit: informationally complete POMs
For the $d = 2$ case of a qubit

$$\rho = \tfrac{1}{2}\left(1 + x\sigma_x + y\sigma_y + z\sigma_z\right), \tag{19}$$

with the $2 \times 2$ matrix of (15) referring to the basis in which the Pauli operators $\sigma_x$, $\sigma_y$, and $\sigma_z$ have their standard form, we have

$$\begin{aligned} x &= \langle \sigma_x \rangle = \sin(2\theta_1) \cos\theta_2 \cos\theta_3, \\ y &= \langle \sigma_y \rangle = \sin(2\theta_1) \cos\theta_2 \sin\theta_3, \\ z &= \langle \sigma_z \rangle = \cos(2\theta_1) \end{aligned} \tag{20}$$

for the expectation values of the Pauli operators. As it should, the sum of their squares

$$x^2 + y^2 + z^2 = 1 - \left[\sin(2\theta_1)\sin\theta_2\right]^2, \tag{21}$$

takes on all values between 0 and 1. For later use, we note that

$$dx\, dy\, dz = d\theta_1 d\theta_2 d\theta_3 \left|\sin(2\theta_1)^3 \sin(2\theta_2)\right|, \tag{22}$$

which exhibits the Jacobian factor that relates the $x, y, z$ parameterization of $\rho$ in (19) to the $\theta$ parameterization of (12) with (16).

We consider two POMs, the four-outcome tetrahedron POM of minimal qubit tomography [8] and the six-outcome Pauli POM that measures in three mutually unbiased bases. For the tetrahedron POM, we have the probabilities

$$\begin{aligned} p_1 &= \frac{1}{4} + \frac{1}{4\sqrt{3}}(x - y - z), \\ p_2 &= \frac{1}{4} + \frac{1}{4\sqrt{3}}(y - z - x), \\ p_3 &= \frac{1}{4} + \frac{1}{4\sqrt{3}}(z - x - y), \\ p_4 &= \frac{1}{4} + \frac{1}{4\sqrt{3}}(x + y + z). \end{aligned} \tag{23}$$

The corresponding constraint factor $w_{\text{cstr}}(p)$ of (I-6) contains $w_{\text{basic}}(p)$ of (I-5) for $K = 4$ and

$$w_{\text{qu}}(p) = \eta\left(\frac{1}{3} - p_1^2 - p_2^2 - p_3^2 - p_4^2\right). \tag{24}$$

By integrating out $p_4$ with the help of the delta function in $w_{\text{basic}}(p)$, we reduce the volume element in the probability space of (I-7) to

$$(dp) \to dp_1 dp_2 dp_3\, \eta\left(\frac{1}{3} - p_1^2 - p_2^2 - p_3^2 - p_4^2\right) \prod_{k=1}^{4} \eta(p_k) \tag{25}$$

with $p_4 = 1 - p_1 - p_2 - p_3$; the $p_k$ values selected by the first step function are non-negative, so that we can omit the $\eta(p_k)$ factors.

Since the Jacobian between $p_1, p_2, p_3$ and $x, y, z$ does not depend on the coordinates, we have

$$dp_1 dp_2 dp_3 \doteq dx\, dy\, dz, \tag{26}$$

and

$$(dp) \doteq dx\, dy\, dz\, \eta(1 - x^2 - y^2 - z^2) \tag{27}$$

follows, where the doubly dotted equal sign says 'essentially equal in the sense of ignoring constant factors and reducing the number of variables such that the remaining ones are independent'. The final step uses (21) and (22) to arrive at

$$(dp) \doteq d\theta_1 d\theta_2 d\theta_3 \left|\sin(2\theta_1)^3 \sin(2\theta_2)\right| \tag{28}$$

for the tetrahedron POM.

For the Pauli POM, we have the six probabilities

$$\left.\begin{matrix} p_1 \\ p_4 \end{matrix}\right\} = \frac{1}{6}(1 \pm x), \quad \left.\begin{matrix} p_2 \\ p_5 \end{matrix}\right\} = \frac{1}{6}(1 \pm y), \quad \left.\begin{matrix} p_3 \\ p_6 \end{matrix}\right\} = \frac{1}{6}(1 \pm z), \tag{29}$$

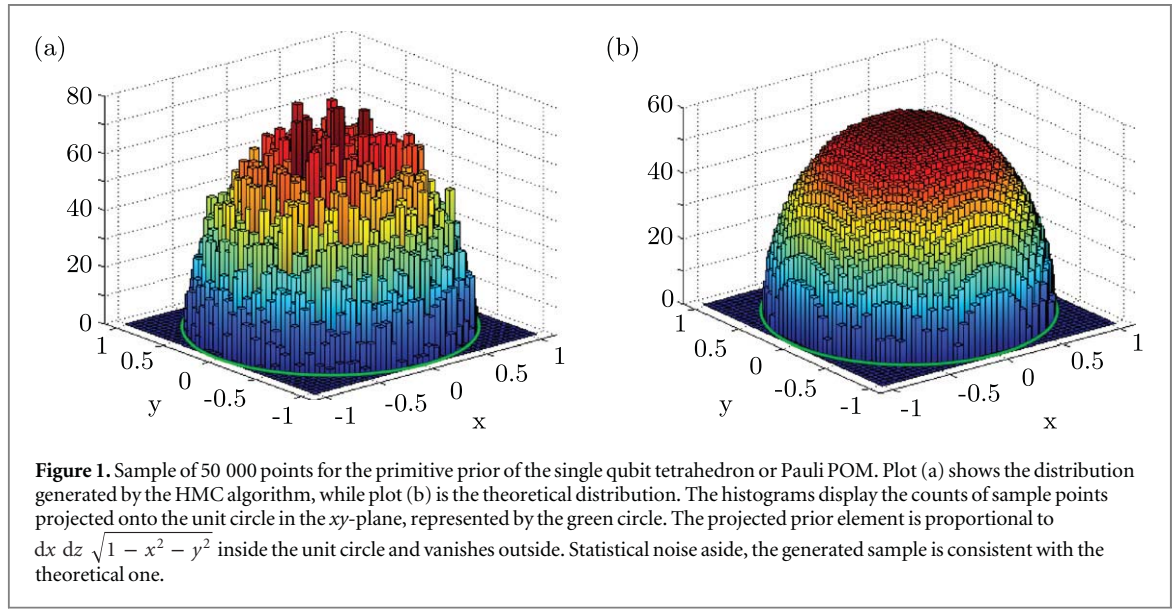

**Figure 1.** Sample of 50 000 points for the primitive prior of the single qubit tetrahedron or Pauli POM. Plot (a) shows the distribution generated by the HMC algorithm, while plot (b) is the theoretical distribution. The histograms display the counts of sample points projected onto the unit circle in the *xy*-plane, represented by the green circle. The projected prior element is proportional to $dx\, dz\, \sqrt{1-x^2-y^2}$ inside the unit circle and vanishes outside. Statistical noise aside, the generated sample is consistent with the theoretical one.

for which

$$w_{\text{basic}}(p) = \prod_{k=1}^{3} \delta\left(p_k + p_{k+3} - \frac{1}{3}\right) \eta(p_k) \eta(p_{k+3}) \tag{30}$$

and

$$w_{\text{qu}}(p) = \eta\left(\frac{1}{9} - \sum_{k=1}^{3} (p_k - p_{k+3})^2\right) \tag{31}$$

are the factors in $w_{\text{cstr}}(p)$. The analog of (25) is

$$(dp) \to dp_1\, dp_2\, dp_3\, \eta\left(1 - \sum_{k=1}^{3}(6p_k - 1)^2\right), \tag{32}$$

and (26)–(28) hold here as well. That is, whether we use the the tetrahedron POM or the Pauli POM, for both the volume element for physical probabilities is that of (28).

Therefore, if we are sampling in accordance with the primitive prior $w_0(p) = 1$, there is no difference between these two POMs. We just have

$$\begin{aligned}
w(\theta) &= \left|\sin(2\theta_1)\right|^3 \sin(2\theta_2), \\
u_1(\theta) &= 6 \cot(2\theta_1), \\
u_2(\theta) &= 2 \cot(2\theta_2), \\
u_3(\theta) &= 0,
\end{aligned} \tag{33}$$

in (4), (8), and (A.2) of the HMC algorithm. The sample of 50 000 points thus produced is reported in figure 1. This sample is a collection of $(x, y, z)$ values inside the Bloch sphere; it is converted into a $p$ sample by (23) or (29), respectively.

Figure 2 shows trajectories of 50 sample points, with consecutive points connected by blue lines, generated using the HMC algorithm, as well as the xMHMC algorithm from I. For the HMC sample, we see that even this short trajectory samples the unit circle rather efficiently, with the points far apart from their predecessors and successors. Such a trajectory is rather different from that of the xMHMC algorithm, where consecutive points are often close together. Indeed, HMC overcomes the problem of the strong autocorrelation that figure I-4 shows. The high acceptance rate of about 95% in the run that produced figures 1 and 2(a) is another advantage of HMC over xMHMC, which had an acceptance rate of about 60% in the run that produced figure 2(b). In higher dimensional cases, the optimal acceptance rates should be about 65% and 25% for the HMC and xMHMC algorithms respectively. And to repeat, since all generated points are physical by construction, there is no need for the CPU-time consuming check of physicality that is necessary in the xMHMC algorithm.

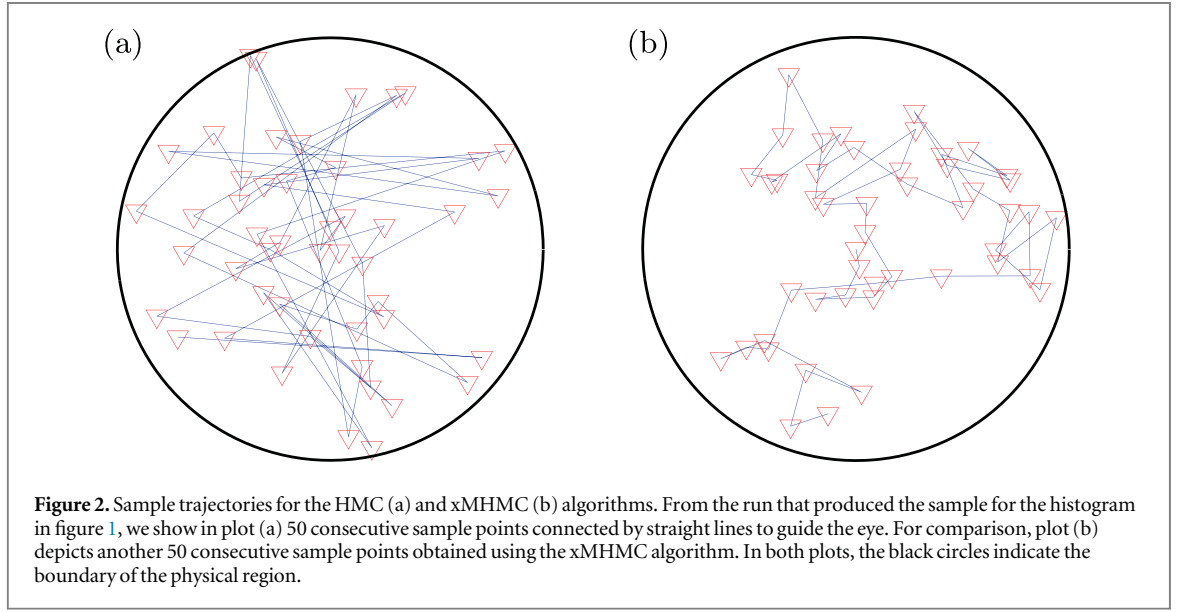

**Figure 2.** Sample trajectories for the HMC (a) and xMHMC (b) algorithms. From the run that produced the sample for the histogram in figure 1, we show in plot (a) 50 consecutive sample points connected by straight lines to guide the eye. For comparison, plot (b) depicts another 50 consecutive sample points obtained using the xMHMC algorithm. In both plots, the black circles indicate the boundary of the physical region.

### 4.2. Qubit: informationally incomplete POMs

We consider two POMs that are not informationally complete, as they provide no information about $z$: the three-outcome trine measurement with the probabilities of (I-4) and the constraint factors

$$w_{\text{basic}}(p) = \delta(p_1 + p_2 + p_3 - 1)\eta(p_1)\eta(p_2)\eta(p_3),$$
$$w_{\text{qu}}(p) = \eta\left(\frac{1}{2} - p_1^2 - p_2^2 - p_3^2\right); \qquad (34)$$

and the four-outcome crosshair POM with the probabilities

$$\left.\begin{array}{c} p_1 \\ p_3 \end{array}\right\} = \frac{1}{4}(1 \pm x), \quad \left.\begin{array}{c} p_2 \\ p_4 \end{array}\right\} = \frac{1}{4}(1 \pm y), \qquad (35)$$

and the constraint factors

$$w_{\text{basic}}(p) = \prod_{k=1,2} \delta\left(p_k + p_{k+2} - \tfrac{1}{2}\right)\eta(p_k)\eta(p_{k+2})$$
$$w_{\text{qu}}(p) = \eta\left(\frac{1}{4} - (p_1 - p_3)^2 - (p_2 - p_4)^2\right). \qquad (36)$$

There are many different reconstruction spaces that we can use; two choices that suggest themselves are the equatorial plane of the Bloch ball

$$\rho = \frac{1}{2}(1 + x\sigma_x + y\sigma_y), \qquad (37)$$

and the upper half of the Bloch sphere

$$\rho = \frac{1}{2}\left(1 + x\sigma_x + y\sigma_y + \sqrt{1 - x^2 - y^2}\,\sigma_z\right), \qquad (38)$$

with $x^2 + y^2 \leqslant 1$ for both. The $\theta_1 = \frac{\pi}{4}$ version of (20) realizes (37)

$$x = \cos\theta_2 \cos\theta_3,$$
$$y = \cos\theta_2 \sin\theta_3,$$
$$z = 0,$$
$$\text{with } dx\,dy \doteq d\theta_2 d\theta_3 \left|\sin(2\theta_2)\right|; \qquad (39)$$

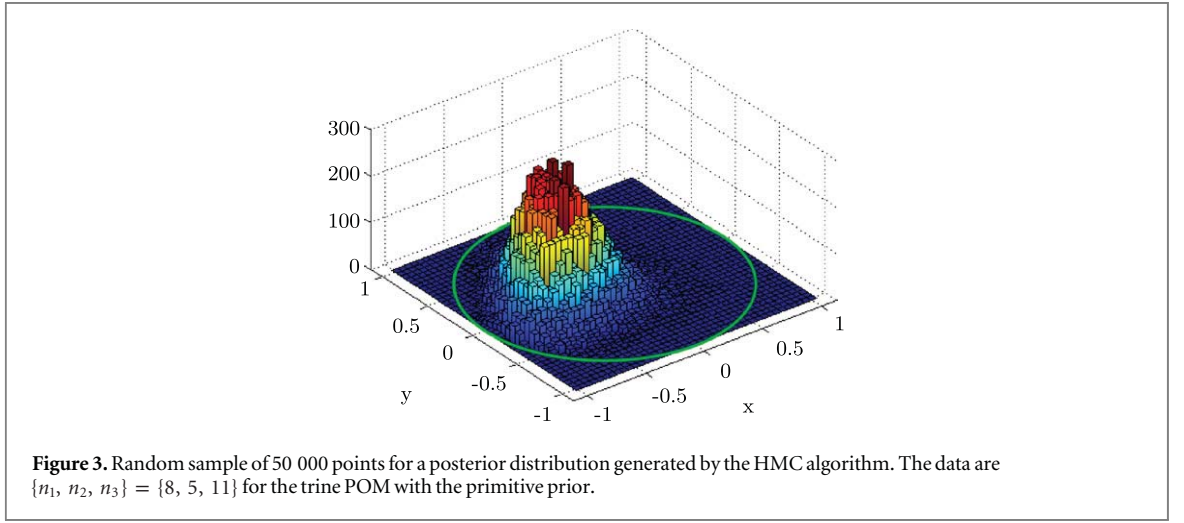

**Figure 3.** Random sample of 50 000 points for a posterior distribution generated by the HMC algorithm. The data are $\{n_1, n_2, n_3\} = \{8, 5, 11\}$ for the trine POM with the primitive prior.

and $\theta_2 = 0$ together with $-\frac{1}{4}\pi \leqslant \theta_1 \leqslant \frac{1}{4}\pi$ fits to (38)

$$x = \sin(2\theta_1)\cos\theta_3,$$
$$y = \sin(2\theta_1)\sin\theta_3,$$
$$z = \cos(2\theta_1),$$
$$\text{with } dx\, dy \doteq d\theta_1 d\theta_3 \left|\sin(4\theta_1)\right|. \tag{40}$$

For the primitive prior density, these give

$$(dp) \doteq \begin{cases} d\theta_2 d\theta_3 \left|\sin(2\theta_2)\right| & \text{for (39)}, \\ d\theta_1 d\theta_3 \left|\sin(4\theta_1)\right| & \text{for (40)} \end{cases} \tag{41}$$

for both the trine POM and the crosshair POM.

Figure 3 shows the histogram of a sample of 50 000 points in accordance with a posterior density. This example is for the trine POM with the primitive prior and the parameterization of (39), and the data are $\{n_1, n_2, n_3\} = \{8, 5, 11\}$. Specifically, we have

$$w(\theta_2, \theta_3) \doteq \left|\sin(2\theta_2)\right| \prod_{k=1}^{3} \left(1 + \cos\theta_2 \cos\theta_3^{(k)}\right)^{n_k} \tag{42}$$

with

$$\theta_3^{(1)} = \theta_3, \quad \theta_3^{(2)} = \theta_3 - \frac{2\pi}{3}, \quad \theta_3^{(3)} = \theta_3 + \frac{2\pi}{3}, \tag{43}$$

and

$$u_2(\theta_2, \theta_3) = 2\cot(2\theta_2) - \sum_{k=1}^{3} n_k \frac{\sin\theta_2 \cos\theta_3^{(k)}}{1 + \cos\theta_2 \cos\theta_3^{(k)}},$$
$$u_3(\theta_2, \theta_3) = -\sum_{k=1}^{3} n_k \frac{\cos\theta_2 \sin\theta_3^{(k)}}{1 + \cos\theta_2 \cos\theta_3^{(k)}}, \tag{44}$$

in (4), (8), and (A.2) of the HMC algorithm. This sample consists of $(x, y)$ pairs in the unit circle and conversion into a sample of *p*s is done with (I-4).

We note that the algorithm that samples in accordance with a posterior specified by the primitive prior and data $D = \{n_1, n_2, n_3\}$ can also sample according to the primitive prior itself, by simply putting $D = \{0, 0, 0\}$. Further, for $D = \{-\frac{1}{2}, -\frac{1}{2}, -\frac{1}{2}\}$ it samples in accordance with the Jeffreys prior, and running the algorithm for $D = \{n_1 - \frac{1}{2}, n_2 - \frac{1}{2}, n_3 - \frac{1}{2}\}$ gives a posterior sample for the Jeffreys prior. Likewise, conjugate priors specified by mock data $\{\nu_1, \nu_2, \nu_3\}$ can be handled by the same algorithm. Analogous remarks apply to POMs with more outcomes.

### 4.3. Qubit pair: double-crosshair POM of BB84

In the BB84 scenario of quantum key distribution, the two parties use two four-outcome crosshair measurements and so have 16 outcomes for the composed POM. In a self-explaining notation that relies on two copies of (35), we denote the joint probabilities by $p_{jk}$ with $j, k = 1, 2, 3, 4$, eight of which are independent. One reconstruction space can be specified by $4 \times 4$ matrices of the form

$$\rho = \rho_0 + \frac{1}{4}q\Sigma,$$

$$\Sigma = \begin{pmatrix} 0 & 0 & 0 & -1 \\ 0 & 0 & 1 & 0 \\ 0 & 1 & 0 & 0 \\ -1 & 0 & 0 & 0 \end{pmatrix},$$

$$\rho_0 = \begin{pmatrix} 4p_{11} & 2(p_{21} - p_{41}) & 2(p_{12} - p_{14}) & p_{\text{even}} \\ 2(p_{21} - p_{41}) & 4p_{31} & p_{\text{even}} & 2(p_{32} - p_{34}) \\ 2(p_{12} - p_{14}) & p_{\text{even}} & 4p_{13} & 2(p_{23} - p_{43}) \\ p_{\text{even}} & 2(p_{32} - p_{34}) & 2(p_{23} - p_{43}) & 4p_{33} \end{pmatrix}, \quad (45)$$

with $p_{\text{even}} = p_{22} - p_{24} - p_{42} + p_{44}$. This unit-trace real symmetric matrix has nine real parameters, eight determined by the probabilities of the double-crosshair POM. While they do not fix the value of the ninth parameter $q = \text{tr}\{\rho \sigma_z^{(1)} \otimes \sigma_z^{(2)}\}$, they determine the range of permissible $q$ values

$$-1 \leqslant q_{\min}(p) \leqslant q \leqslant q_{\max}(p) \leqslant 1. \quad (46)$$

The $p$ dependent bounds $q_{\min/\max}(p)$ can be found by requiring $\rho \geqslant 0$, as will be discussed shortly. By choosing a specific $q$ value—perhaps the largest permissible value $q_{\max}(p)$, or the value closest to 0—we get a definite reconstruction space. This is, however, awkward if we want to sample by the HMC algorithm.

In order to find $q_{\min/\max}(p)$, we demand that all the eigenvalues of $\rho$ must be non-negative. The determinant of $\rho$ is a forth-order polynomial in $q$

$$\det\{\rho(q)\} = \left(\frac{q}{4}\right)^4 - \frac{1}{2}\text{tr}\{(\rho_0\Sigma)^2\}\left(\frac{q}{4}\right)^2 + \text{tr}\{(\rho_0^2 - \rho_0^3)\Sigma\}\left(\frac{q}{4}\right) + \det\{\rho_0\}. \quad (47)$$

Being the product of the eigenvalues, the roots of the determinant are the $q$ values for which one of the eigenvalues of $\rho(q)$ equals to zero. Hence, the range of physical $q$ is bounded by roots of (47). The positive coefficient of the $q^4$ term implies that the determinant is positive at regions of large $|q|$. Given that the roots are the $q$ values where the determinant changes signs, the only other region where the determinant can be positive is between the second and third roots. From (45), it is clear that for large $|q|$, $\rho$ will have two positive and two negative eigenvalues. Hence, the only permissible region is the region bounded by the second and third roots of (47). Finding these roots therefore gives us $q_{\min}$ and $q_{\max}$.

Since we get the matrix $\rho$ in (45) from the $4 \times 4$ matrix in (15) by setting $E_{10} = \cdots = E_{15} = 1$, which leaves us with the nine angle parameters $\theta_1, \theta_2, \ldots, \theta_9$, we supplement the eight-dimensional probability element $(dp)w(p)$ with a factor for $q$

$$(dp)w(p) \to (dp)dq\, w(p, q) = \frac{(dp)dq\, w(p)}{q_{\max}(p) - q_{\min}(p)},$$

$$w(p) = \int dq\, w(p, q). \quad (48)$$

and sample from the nine-dimensional target density $w(p, q)$. Upon ignoring the $q$ values of the sample points $(p^{(j)}, q^{(j)})$, we get the wanted sample of probabilities $p^{(j)}$. Ideally, we would like to sample from $w(p, q)$ by means of HMC sampling. However, due to $q_{\min/\max}(p)$ taking very complicated forms, it would not be very inconvenient to compute the derivatives of the potential, which is required by HMC. What can be done more simply is to perform HMC to sample from the distribution $w^*(p, q) = w(p)$, followed by importance sampling (as described in [1]) that assigns each point the weight $(q_{\max}(p) - q_{\min}(p))^{-1}$, effectively leaving us with the desired distibution $w(p, q) = w(p)$.

We proceed to analyze the distribution of the Clauser–Horne–Shimony–Holt (CHSH) quantity $S$ obtained using our sampling method. Here, $S$ is defined as [9, 10]

$$S = E(A_1, B_1) + E(A_2, B_1) + E(A_1, B_2) - E(A_2, B_2), \quad (49)$$

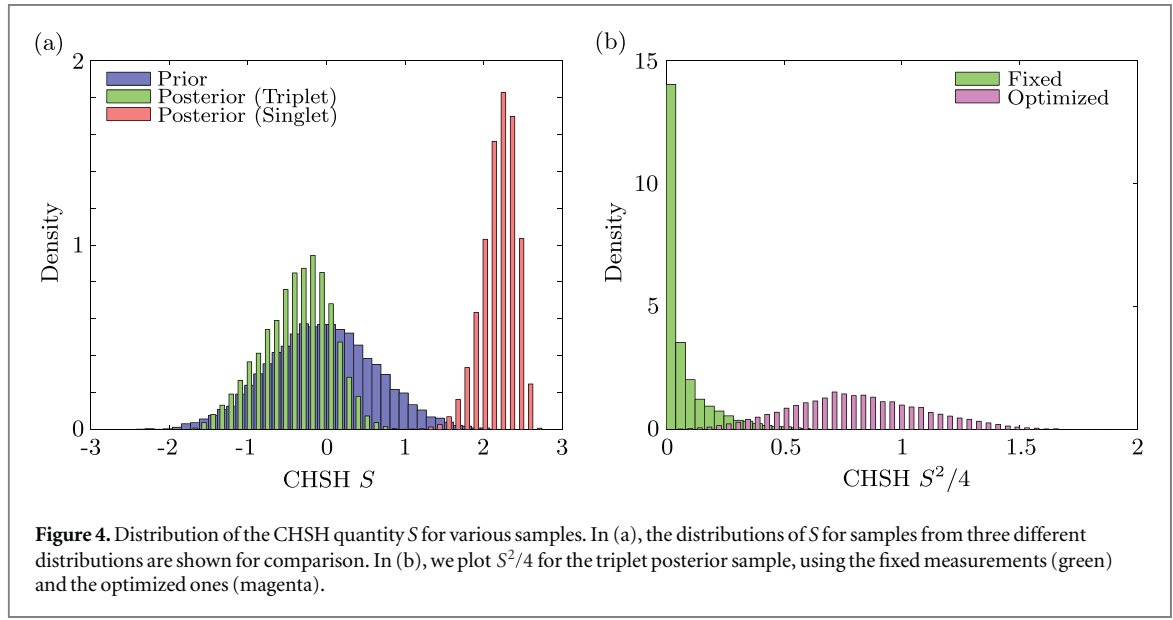

**Figure 4.** Distribution of the CHSH quantity *S* for various samples. In (a), the distributions of *S* for samples from three different distributions are shown for comparison. In (b), we plot $S^2/4$ for the triplet posterior sample, using the fixed measurements (green) and the optimized ones (magenta).

where $A_1$ and $A_2$ are observables on the first qubit, while $B_1$ and $B_2$ are observables on the second qubit, each having eigenvalues of $\pm 1$. For a state $\rho$, $E(A, B)$ is given by

$$E(A, B) = \langle A \otimes B \rangle = \mathrm{tr}\{\rho A \otimes B\}. \tag{50}$$

The CHSH quantity ranges from $-2\sqrt{2}$ to $2\sqrt{2}$, with values $|S| > 2$ being evidence of quantum entanglement between the qubits. For this example, we restrict measurements $A_i$ and $B_j$ to the form

$$A_i = \sigma_x \cos\phi_i + \sigma_y \sin\phi_i,$$
$$B_j = \sigma_x \cos\psi_j + \sigma_y \sin\psi_j. \tag{51}$$

In general, *S* depends on $\phi_i$ and $\psi_j$. For any state $\rho$, *S* can be maximized by optimizing $\phi_i$ and $\psi_j$, which gives

$$S = 2\sqrt{\sum_{i,j=x,y} \langle \sigma_i \otimes \sigma_j \rangle^2}. \tag{52}$$

We begin by fixing $\phi_1 = 0$, $\phi_2 = \pi/2$, $\psi_1 = 5\pi/4$, and $\psi_2 = 3\pi/4$. With this setting, *S* can be expressed in terms of our detector probabilities

$$S = 8\sqrt{2}\left(p_{12} + p_{13} + p_{21} + p_{23} + p_{31} + p_{32} - 2p_{22}\right) - 2\sqrt{2}. \tag{53}$$

In figure 4(a), the resulting distribution of the CHSH quantity is shown for three different samples. The first, in blue, is drawn from the primitive prior. The next, in green, is drawn from the posterior obtained from data that corresponds to the triplet state $\frac{1}{\sqrt{2}}(|10\rangle + |01\rangle)$ with noise. Lastly, the red distribution is drawn from the posterior density obtained from data corresponding to the singlet state $\frac{1}{\sqrt{2}}(|10\rangle - |01\rangle)$. Each sample contains 50 000 sample points, and the data sets used for the posterior densities are made up of 64 measured qubit pairs each. It can be seen that although the triplet state has as much entanglement as the singlet state, the CHSH quantities for the triplet sample are much closer to 0 for the unoptimized measurements. In figure 4(b), we compare the quantity $S^2/4$ obtained by using the fixed measurement (in green) against that obtained from the maximized CHSH quantity (in magenta) for the sample drawn from the posterior density of the triplet state with noise. This quantity $S^2/4$ takes values between 0 and 2, with values larger than 1 being evidence for quantum entanglement. For the fixed measurement of (53), the values that exceed $S^2/4 = 1$ are extremely rare. By contrast, there is a large fraction of values with $S^2/4 > 1$ for the optimized *S* value of (52).

In closing, we note another use of (45). It can provide a physicality check for candidate probabilities $p_{jk}$ that is quite different from, and much simpler than, the algorithm discussed in appendix A in [1]. Given $p_{jk}$s that obey all the basic constraints, they are physical if there is a *q* value for which $\rho \geq 0$; otherwise, they are not physical.

## 5. Conclusion

We have adapted the HMC sampling algorithm to the problem of sampling quantum state spaces, where the constraints that result from the positivity of the statistical operator must be obeyed throughout. To ensure this, we use a systematic parameterization of the statistical operator without any superfluous parameters. This can always be done for informationally complete measurements, and is also possible when the measurement is not informationally complete as long as a suitable parameterization of the relevant subspace of the state space is at hand. In all other cases, one must resort to other sampling algorithms, such as the ones we discuss in [1].

## Acknowledgments

We thank Michael Evans and Yong Siah Teo for stimulating discussions. This work is funded by the Singapore Ministry of Education (partly through the Academic Research Fund Tier 3 MOE2012-T3-1-009) and the National Research Foundation of Singapore.

## Appendix. The leapfrog method

The leapfrog method is a split-operator method much like those based on the Trotter–Suzuki formula (see, e.g., [11]). For a small time increment $\tau$, we advance $(\theta(t), \vartheta(t))$ to $(\theta(t+\tau), \vartheta(t+\tau))$ in three steps: by letting only the kinetic energy $\frac{1}{2}\sum_s \vartheta_s^2$ govern the evolution for duration $\frac{1}{2}\tau$ in the first and third steps, whereas only the potential energy $-\log w(\theta)$ is relevant in the intermediate second step of duration $\tau$. In total, this amounts to

$$(\theta(t), \vartheta(t)) \rightarrow (\theta(t+\tau), \vartheta(t+\tau)),$$
$$\text{with } \theta(t+\tau) \cong \theta(t) + \tau\vartheta(t) + \tfrac{1}{2}\tau^2 u\left(\theta(t) + \tfrac{1}{2}\tau\vartheta(t)\right),$$
$$\vartheta(t+\tau) \cong \vartheta(t) + \tau u\left(\theta(t) + \tfrac{1}{2}\tau\vartheta(t)\right), \quad (A.1)$$

where the right-hand sides are correct to order $\tau^2$ and have an error $\propto \tau^3$.

We break up the total duration $T$ into $L$ time intervals $\tau = T/L$ so that $L$ leapfrog jumps accomplish HMC3 with a discretization error $\propto L\tau^3 = T^3/L^2$. The two adjacent kinetic-energy $\frac{1}{2}\tau$-periods of subsequent jumps are conveniently combined into a single period of a full $\tau$. Accordingly, we find the final $(\theta^\star, \vartheta^\star)$ pair from the initial $(\theta, \vartheta)$ pair by this procedure:

**LF1** Set $(\theta_1, \vartheta_1) = (\theta + \tfrac{1}{2}\tau\vartheta, \vartheta)$.
**LF2** For $j = 1, 2, 3, ..., 2L-1$, set

$$(\theta_{j+1}, \vartheta_{j+1}) = \begin{cases} (\theta_j, \vartheta_j + \tau u(\theta_j)), & \text{for } j \text{ odd}, \\ (\theta_j + \tau\vartheta_j, \vartheta_j), & \text{for } j \text{ even}. \end{cases} \quad (A.2)$$

**LF3** Set $(\theta^\star, \vartheta^\star) = (\theta_{2L} + \tfrac{1}{2}\tau\vartheta_{2L}, -\vartheta_{2L})$.

We note an important property of the leapfrog method: the individual steps in LF1–3 are shearing transformations that preserve phase-space volumes, so that the approximate solution of (8) is volume-preserving just like the exact one (Liouville's theorem).

According to Neal [3], there is a trade-off between accuracy in the propagation and CPU time consumed. A good choice of $L$ is such that the acceptance rate (10) is about 65% for high-dimensional problems. Further, if one observes slow convergence, the likely cause is nonergodicity, which can be cured by randomly choosing $\tau$ and $L$ from fairly small intervals [12].